\documentclass[10pt,twocolumn]{article}

\usepackage{lipsum} 		
\usepackage{blindtext} 		

\usepackage[superscript, biblabel]{cite}  	

\makeatletter
\renewcommand\@biblabel[1]{#1.}
\makeatother

\usepackage{etoolbox}
\patchcmd{\thebibliography}{\section*{\refname}}{}{}{}

\usepackage[margin=1.0in,hmarginratio=1:1,top=32mm,columnsep=15pt]{geometry}

\usepackage{color}				
\definecolor{dark-gray}{gray}{0.1}	
\usepackage{times}				
\usepackage{microtype} 			
\usepackage[english]{babel} 		

\usepackage[hang, font={footnotesize}, labelfont={bf,up}, textfont={sf,up}]{caption} 
\usepackage{booktabs} 	
\usepackage{mwe}  		

\usepackage{enumitem} 		
\setlist[itemize]{noitemsep} 	

\usepackage{abstract} 
\AtBeginDocument{}  		

\usepackage{titlesec} 			
\renewcommand\thesection{\Roman{section}.} 		  		
\renewcommand\thesubsection{\thesection\Alph{subsection}.} 	
\renewcommand\thesubsubsection{\thesubsection\arabic{subsubsection}.} 
\titleformat{\section}[block]{\normalfont\sffamily\bfseries}{\thesection}{1em}{\MakeUppercase}{} 	
\titleformat{\subsection}[block]{\normalfont\sffamily\bfseries}{\thesubsection}{1em}{}{}  
\titleformat{\subsubsection}[block]{\normalfont\sffamily\bfseries}{\thesubsubsection}{1em}{}{}  
\titlespacing*{\section}{0.0em}{1em}{0.25em}		
\titlespacing*{\subsection}{0.0em}{1em}{0.25em}	
\usepackage{indentfirst}						

\usepackage{tocloft}
\usepackage{fancyhdr} 

\fancypagestyle{plain}{
\fancyhf{}
\lhead{\color{dark-gray}\textit{}} 
\rhead{ {\it Submitted to ANS/NT (2021). 'LA-UR-21-21605 DRAFT} }
}

\usepackage{titling} 		

\usepackage{hyperref} 	
\hypersetup{
	backref=true,       
    	pagebackref=true,               
    	hyperindex=true,                
    	colorlinks=true,                
    	breaklinks=true,                
    	urlcolor= black,                
    	linkcolor=black,                
    	bookmarks=true,                 
    	bookmarksopen=false,
    	filecolor=black,
    	citecolor=blue
}

\pretitle{\begin{center}\large\bfseries} 	
\posttitle{\end{center}} 				
\title{\vspace{-0.3in} \sffamily{Foreword: \\
Manhattan Project Nuclear Science \& Technology Developments at Los Alamos } }	
\author{%
\normalsize Mark B. Chadwick\thanks{corresponding author: mbchadwick@lanl.gov}\\[-0.5ex] 
\normalsize Los Alamos National Laboratory \\[-0.5ex] 
\normalsize Los Alamos, NM 87545
}
\date{ } 

\begin{document}

\maketitle	

 The year 2020 marked the 75th anniversary of the Trinity experiment, the world's first nuclear explosion, on July 16, 1945, near Alamogordo, New Mexico. Trinity was a vital proof step toward the culmination of the Manhattan Project and the end of  World War II. The technical accomplishments made by scientists and engineers from the US, UK, and Canada (some originating in Germany, Hungary, Italy, France, and other countries) were recognized by many events in 2020, including  a visit to New Mexico's Los Alamos National Laboratory by Department of Energy (DOE) National Nuclear Security Administration (NNSA) dignitaries, historical documentaries \cite{Sarrao:2020,Carr:2020}, and the publication of an excellent book, {\it Trinity,} by physicist Frank Close \cite{Close:2019}. The importance of Trinity as a foundational accomplishment for the broad nuclear science and engineering community is clear; indeed, New Mexico’s chapter of the American Nuclear Society (ANS) is referred to as the “Trinity Section.” The events surrounding Trinity have even entered into high culture with recent performances of John Adams's opera {\it Doctor Atomic} in San Francisco, Amsterdam, Chicago, New York, and Santa Fe.

At a high-powered meeting in May 1945, with Vannebar Bush, Gen. George Marshall, Gen. Leslie Groves, Arthur Compton, James Conant, Robert Oppenheimer, Enrico Fermi, and Ernest Lawrence in attendance, Secretary of War Harry L. Stimson spoke the prescient (if somewhat grandiose) words: {\it ``This project [Manhattan Project] should not be considered simply in terms of military weapons, but as a new relationship of man to the universe. This discovery might be compared to the discoveries of the Copernican theory and of the laws of gravity, but far more important than these in its effect on the lives of men. While the advances in the field to date had been fostered by the needs of war, it was important to realize that the implications of the project went far beyond the needs of the present war. It must be controlled if possible to make it an assurance of future peace rather than menace to civilization'' }\cite{Stimson:1945}.

Our universities' science and engineering luminaries, together with their best graduate students, came together at various Manhattan Project locations across the country under intense pressure and worked at a frenetic wartime pace to successfully develop a workable atomic bomb. This gathering of scientific and technical excellence was a unique event in our history, and their collective effort led to a remarkable outpouring of scientific creativity in nuclear and material sciences and in hydrodynamics and neutronics computations, and led to the creation of what would become today’s US DOE national laboratories. The historic effort was facilitated in large part by Oppenheimer's superb (and, at the time, unproven) leadership. The Manhattan Project, together with the MIT Radiation Laboratory's wartime work on radar and E.O. Lawrence's accelerator research, represents the beginnings of big science, bringing thousands of researchers together to solve problems, a model that has since proved so effective for the scientific community in the field of particle physics, sequencing of the human genome, and the discovery of gravitational waves. 

To recognize the Trinity anniversary, this special issue of {\it Nuclear Technology} focuses on aspects of the science and engineering breakthroughs made during the Manhattan Project at Los Alamos (then called Project Y), 1943–1945. Both technical papers and technical history papers are included. Their goal is to clarify the nature of the breakthroughs made, correct previous misunderstandings in the open literature, illuminate fascinating aspects of the underlying research, and illustrate how science from 75 years ago has proven foundational for the peaceful use of nuclear energy and today’s nuclear technology. These papers benefit from the authors’ access to extensive material in our Los Alamos National Security Research Center (NSRC) archives, much of which has not been available to broader audiences. 

The idea of this special issue grew out of a seminar series that I hosted at Los Alamos in the summer of 2020. Afterward, I challenged our staff to write up their work as technical papers to be published in an internal Los Alamos classified research journal. The invitation was extended to colleagues at the Lawrence Livermore National Laboratory, Sandia National Laboratories, and the Atomic Weapons Establishment (AWE) at Aldermaston in the UK (appropriate given the large British contribution to the Manhattan Project). The authors overcame substantial challenges in completing the papers during the COVID-19 pandemic, not least of which was the limited access to our onsite archives. Despite this, our project led to the fascinating suite of papers presented here. Subsequent discussions with the American Nuclear Society resulted in the decision to make this scholarship available to a broader readership by publishing the unclassified papers in this special issue of {\it Nuclear Technology} with open access.

Below, I summarize some of the key insights to be found in this collection of papers.

{\bf Nuclear science and technology.} Ref.\cite{Chadwick:2021b} documents the neutron cross sections measured with increasing accuracy during the Manhattan Project. Accurate neutron cross sections were needed to determine critical masses and, therefore, the quantities of $^{235}$U and $^{239}$Pu required from Oak Ridge and Hanford.  In short order, four university accelerators were disassembled and reassembled at Los Alamos, and methods were 
established to make measurements on extremely small samples owing to the initial lack of availability of enriched $^{235}$U and plutonium. In just two years, advances in experimental methods led to measured nuclear data that are surprisingly close to today’s best values in our Evaluated Nuclear Data Files (ENDF). Many of the key original papers and numerical values have now been archived through a collaboration with the International Atomic Energy Agency (IAEA) and Brookhaven National Laboratory in the internationally available Experimental Nuclear Reaction Data (EXFOR) database. Also, an early, little-known British paper is transcribed \cite{Chadwick:2021b} : Bretscher’s 1940 description of the usefulness of plutonium, written around the time that the element was discovered. 

Papers on the first fast critical assemblies (Hutchinson et al. \cite{Hutchinson:2021}) and pulsed and solution assembly experiments (Kimpland et al. \cite{Kimpland:2021}) provide details about how critical masses were determined and how they influenced subsequent research across the world on nuclear criticality and criticality safety. The Los Alamos ``water boiler'' assembly was the world's third reactor to become operational (in 1944, after Chicago's CP-1 and Oak Ridge's X-10 piles), the first to use a solution, and the first to use enriched uranium fuel. During the Manhattan Project, bare critical masses were not measured directly–-there was insufficient time or material; instead, the wartime measurements focused on reflected assemblies and subcritical measurements from which bare critical masses could be estimated with extrapolative calculations. Sood et al. \cite{Sood:2021} describe the evolution of neutronics calculational capabilities from early neutron diffusion work to subsequent refinements by Serber and Wilson (a British researcher) and the postwar innovations of $S_n$ deterministic and Monte Carlo neutron-transport simulations.

Andrews, Andrews, and Mason \cite{Andrews:2021b} describe the Canadian work at the Montreal Laboratory and Chalk River and the essential role Canada played in supplying nuclear materials for the Manhattan Project. The authors also tell of the contributions of the talented Canadians who came to work on the Manhattan Project in the US. The Montreal Laboratory's work was focused on neutronic criticality theory and heavy-water-moderated reactor experimentation––research that proved to be important for postwar CANDU reactor development.

{\bf Hydrodynamics.} Morgan and Archer \cite{Morgan:2021} describe Los Alamos' Theoretical Division’s Lagrangian hydrodynamic shock calculations, implemented on IBM punched-card machines. Their paper presents the algorithmic advances made during the Manhattan Project by von Neumann that led to the late-1940s formulation of computational fluid dynamics by von Neumann and Richtmyer that is today the basis of simulations of everything from climate change to nuclear reactor design. But Morgan and Archer also illuminate the less appreciated, but very influential, roles of Peierls and Skyrme. The authors show that the first usage of ``artificial viscosity," a concept central to computational hydrodynamics, appears to originate with Peierls in 1944. Skyrme is well known to nuclear and particle physicists, but few know of his shock-physics research. Indeed, it was the  expertise in shock physics that was especially sought after by Oppenheimer and  Bethe in the British Mission that brought two dozen British scientists to New Mexico in 1944. Other papers describe the history of the Los Alamos computing facility, focusing first on the “human computers” (Lewis \cite{Lewis:2021}) and then the IBM punched-card computations (Archer \cite{Archer:2021}) needed for hydrodynamics and neutronics.
 
{\bf High explosives.} The implosion design of the Fat Man atomic bomb relied on precision-engineered high explosives (HE) to symmetrically compress a solid ball of plutonium. Brown and Borovina \cite{Brown:2021} describe this HE work, its subsequent impact on broader shaped-charge technology, and its use in mining, oil recovery, and even SpaceX multistage rocket separation. AWE's Moore \cite{Moore:2021} describes pioneering British work on explosive shaped charges that influenced von Neumann, Neddermeyer, and Tuck’s HE lens design. Indeed, it was new to me that both types of explosives used in the Trinity explosive lens system––``Comp B’’ (a mixture of RDX and TNT) and Baratol––had their origins in earlier British defense research on HE formulation. Morgan’s paper \cite{Morgan:2021b} describes the Jumbo steel vessel designed, if the Trinity test should have failed, to contain the Trinity gadget and conventional explosion and allow recovery of the precious plutonium. In the end, Jumbo was not used for Trinity, but the experience gained was valuable for later containment-vessel work and reactor engineering. 

{\bf Plutonium materials \& metallurgy.} Martz, Freibert, and Clark \cite{Martz:2021} trace the exciting process through which the new element plutonium was discovered in 1940 at University of California–Berkeley. The first plutonium was characterized there and at the Chicago’s Metallurgical Laboratory, before US research efforts were consolidated at Los Alamos in 1943. Particularly interesting is the early confusion caused by the widely varying density measurements and the subsequent discovery of the many complex phases of plutonium. This work collects the historical records and reconstructs the history of the rapidly advancing field of plutonium metallurgy and chemistry. The authors show that the idea of using gallium as an alloying agent to stabilize the malleable $\delta$-phase of plutonium was first raised only a few months before the Trinity test, a reflection of the intense pace of the project. They also describe postwar work on the world’s first fast reactor, Clementine, which operated in a canyon, no less, at Los Alamos. Another paper by Crockett and Freibert \cite{Crockett:2021} describes the rapid wartime expansion of experimental and theoretical work on the equation of state (EOS). They describe the foundational EOS research needed to understand the hydrodynamic behavior of plutonium, uranium, and other materials. Los Alamos remains the DOE's center of excellence for plutonium research, signified in a joint effort with the NNSA and AWE by the seven-volume {\it Plutonium Handbook}  \cite{Clark:2019}, published by the ANS in 2019. The wartime plutonium work, unlike the other areas described above, did not benefit from earlier British research, except $\ldots$ Los Alamos’s leading metallurgist, Cyril Stanley Smith, was a naturalized US citizen who had emigrated from England!

{\bf Nuclear energy and yield.} Hanson and Oldham \cite{Hanson:2021} review the foundational radiochemistry methods developed to measure the yield of Trinity and how the techniques evolved in subsequent years. Mercer et al. \cite{Mercer:2021} describe recent measurements of radionuclides in trinitite rock from Alamogordo. Their paper describes both traditional radiation-detection methods that are used in the training of IAEA inspectors at Los Alamos as well as the novel decay energy spectroscopy method.

In the years following the war, a yield of 21 kilotons (kt) for Trinity was announced by the DOE. This was higher than many of the Los Alamos scientists had originally expected––the Fat Man “gadget” overperformed. Now, Selby et al.'s paper \cite{Selby:2021} shares our latest assessment that Trinity’s yield was higher still, 24.8$\pm$2\,kt. The new value comes from advances in precision mass spectrometry and related calculational methods that allow us to reanalyze the 75-year-old debris, measuring the stable nuclides into which the original radionuclides have decayed. Other papers examine early prompt assessments of Trinity's yield: Katz \cite{Katz:2021} seeks to understand how Fermi might have determined the yield when he observed the blast wave’s impact on small pieces of falling paper; Baty and Ramsey \cite{Baty:2021} revisit G. I. Taylor’s determination of the yield from the growth of the fireball. Using Lie group symmetry techniques, they rederive Taylor's two-fifths law relating a blast wave's position, time, and explosive energy. 

During the Manhattan Project, Bethe and Feynman developed an analytic formula to predict the yield of a fission explosion from some elegant considerations. This work has had an enduring influence over the past 75 years; our classified research journal on Trinity has no fewer than five papers on different aspects of the formula––three from Los Alamos, one from Livermore, and one from AWE. 
A short paper by Lestone  and Rosen (Livermore) \cite{Lestone:2021} 
describes, for the first time, the formula and its relationship to earlier wartime British work by Frisch and Peierls and by Pryce and  Dirac. (Wigner, who was Dirac’s brother-in-law, said that ``Feynman is a second Dirac, only this time human” \cite{Farmelo:2011}!). 

{\bf Technical history.} This special issue includes a number of technical history papers, including those mentioned above on the beginnings of computing.  Another \cite{Chadwick:2021c} discusses the origins of the nuclear-core design for the Trinity implosion and presents details of the invention patent by Christy and Peierls that is held in our NSRC archives, resolving longstanding disputes regarding who originated the idea (it {\it was} Christy). Moore gives \cite{Moore:2021a} an introduction to Peierls’ fascinating 1945 summary of the British contributions to the ``Tube Alloys’’ project, the codename for the British work before it transferred to Los Alamos in 1943--1944. The author provides Peierls' summary in full––with Sir James Chadwick's marginal notes. Moore's useful introduction and footnotes shine light on the activities of the time and the progress by the British researchers toward establishing the feasibility of an atomic bomb.

{\bf Trinity and its impacts.} I will end where we will begin. The  first paper in our special issue is by Carr \cite{Carr:2021b}, our Laboratory historian. He gives an overview of the Trinity test and its impact, explains why Oppenheimer chose the name, and provides additional context for the papers that follow. Carr also includes information from our NSRC archives related to the effects of the fallout from the test; also see the useful 2020 special issue in {\it Health Physics} \cite{Simon:2020} published by the National Cancer Institute (NCI). The NCI study was mandated by Congress and took seven years to complete, producing six impressive peer-reviewed articles. 

Our goal for this collection of papers is that they reflect Robert Wilson’s 1947 words in Los Alamos report LA-1009: “{\it The work reported was for a specific military purpose. It will be gratifying to all those who participated in the work when it takes its more proper place as a contribution to the general structure of scientific knowledge}.'' I can speak for all the authors when I say that we had fun writing these papers and that we learned many new things in the process, and I trust that this collection is indeed a contribution to both the history of science and to the advancement of science. It is clear how innovations in nuclear science and technology continue to benefit from cross-fertilization between national security-driven  and civilian programs. Furthermore, it is evident that  foreign-born scientists have contributed substantially to US research advances. 

{\bf Acknowledgments.} I would like to acknowledge staff who played an exceptional role in writing numerous articles and in providing extensive valuable ``internal’’ review feedback to our authors: Bill Archer and Craig Carmer (Los Alamos); Mordy Rosen (Livermore); Richard Moore and Peter Adsley (AWE); and Jonathan Katz (Washington University and Los Alamos). Furthermore, {\it Critical Assembly}, the excellent book on the Manhattan Project, written by Los Alamos historians in the early 1990s \cite{Hoddeson:1993}, provided an important foundation for our research, and usefully identifies primary source archival material held in our National Security Research Center (NSRC). I am also grateful to Tom Kunkle who provided pointers to many valuable historical documents for our project. A paper that Kunkle is presently writing describes the German origins of implosion concepts, starting around 1940 and including the role of Guderley's 1942 hydrodynamics paper. Kunkle addresses the extent to which it was known or commissioned by Schumann for wartime German atomic energy research, and indeed whether any of this was known by US Manhattan Project 
researchers at the time as Lowell Wood has suggested \cite{Wood:2020b}––stay tuned!

This work was supported by the US Department of Energy through the Los Alamos National Laboratory. Los Alamos
National Laboratory is operated by Triad National Security, LLC, for the NNSA of the US DOE under Contract No.
89233218CNA000001.


\vspace{0.25in}
\noindent\rule{0.35\textwidth}{.4pt}


%

\bibliographystyle{ans_js}   
  \small\bibliography{bibliography.bib}  

\end{document}